# Effects of Lithographic Stitching Errors on the Performance of Waveguide Bragg Gratings

Steve Zamek*, Mercedeh Khajavikhan, Dawn T.H. Tan, Maurice Ayache, Boris Slutsky, and Yeshaiahu Fainman

*Department of Electrical and Computer Engineering, University of California, San Diego, 9500 Gilman Drive, La Jolla, CA 92093-0407*
*\*szamek@ucsd.edu*

**Abstract:** We investigate the performance of waveguide Bragg gratings as a function of imperfections introduced in the fabrication process. Effects of stitching errors introduced in the electron-beam and UV-lithography are discussed in details.

## 1. Introduction.

Chip-scale photonic circuits promise to revolutionize applications ranging from on-chip interconnects to long-haul optical cross-connects. Devices based on Bragg gratings have been extensively used in lasers, amplifiers, channel selectors, wavelength filters, routers, and numerous biomedical applications. Fabrication techniques and the associated imperfections were extensively studied during the past decades for fiber Bragg gratings [1]. Some efforts were undertaken to address these issues in chip-scale waveguide Bragg gratings [2-4]. The quest for repeatable large-volume fabrication of such gratings requires models that can provide better understanding of the sources and the effects of fabrication imperfections in those structures.

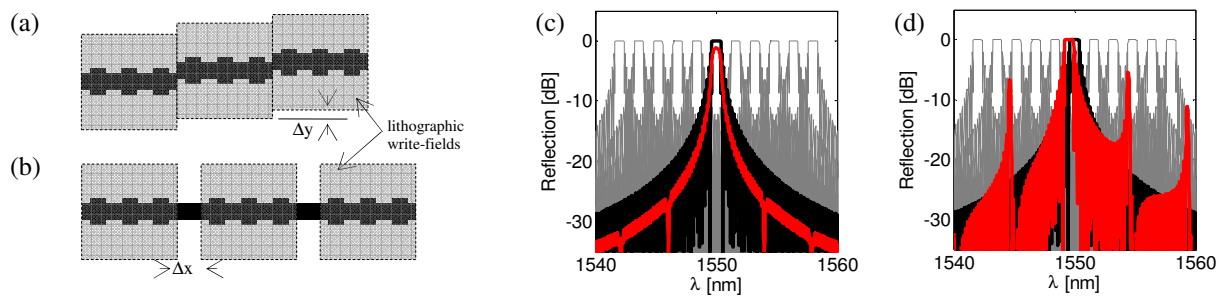

Figure 1. Illustration of the origins and effects of stitching errors. (a) and (b) show lateral (Δy) and longitudinal offsets (Δx), respectively. Their effects are shown in (c) and (d), respectively for 30 stitches with systematic offsets of 30nm each. On both figures the response of an ideal filter (no stitches) is shown in black, the one of a filter with the stitches is shown in red, and the other channels in the grid are in gray.

A Bragg grating with critical dimension of the order of 100 nm requires fabrication precision of about 10 nm. On the other hand, filter bandwidths of 0.2 nm, typical in dense WDM systems, require structures with lengths exceeding 5 mm. The combination of high resolution and large dimensions of the structure impose a fabrication challenge. Fabrication can be done by either direct electron beam (e-beam) writing, or by UV lithography, utilizing an e-beam written mask. In both cases, errors in the stitching of multiple e-beam write-fields cause lateral and longitudinal offsets, as shown in Fig 1a and 1b, respectively. In this work we propose a model and apply it to study the effects of stitching on waveguide Bragg gratings. Although the model allows analysis of both systematic and random offsets, we focus on the former. In the existing e-beam systems systematic offsets are typically larger than the random ones, and their effect is more severe due to the coherent reflections within the Bragg resonator.

## 2. Theoretical Model

The elements of our structure are sections of Bragg grating, step discontinuities introduced by the lateral stitch offsets (see Fig 1a), and straight waveguides created by the longitudinal stitch offsets (see Fig 1b). Each element is represented by a transmission matrix [5]:

$$T = \begin{bmatrix} (1-\alpha)t^{*-1} & -r^*t^{*-1} \\ -rt^{-1} & t^{-1} \end{bmatrix} \quad (1)$$

For a section of waveguide Bragg grating, the coefficients *r* and *t* in Eq. 1 can be found from Coupled Mode Theory [6]. For longitudinal offsets, T is simply a matrix describing a straight waveguide, with $r=\alpha=0$ and $t=exp(-j\beta\Delta x)$, $\beta$ is the propagation constant for the mode of interest, and $\Delta x$ is the longitudinal displacement (see Fig 1). Lateral displacements are step discontinuities, whose T-matrix is uniquely defined by the amount of loss at each discontinuity [7].

To investigate the effect of field offsets, we consider a Bragg grating with N stitches. If *m*-th stitch is described by a matrix $T_{S,m}$, the entire cascade is described by $T_{TOT}=T_B \cdot T_{S,1} \cdot T_B \cdot \ldots \cdot T_{S,N}$ where $T_B$ corresponds to a Bragg grating that fits into a single write-field of the e-beam writer. Transmission and reflection coefficients can be obtained from $T_{TOT}$:

$$t_{TOT} = (T_{TOT})_{2,2}^{-1}; \quad r_{TOT} = -(T_{TOT})_{1,2}(T_{TOT})_{2,2}^{-1} \qquad (2)$$

Here the indices *i,j* designate *i*-th row and *j*-th column of the matrix. Once all the displacements are known, the T-matrices can be calculated for all elements of the structure, the total T-matrix is obtained, and the transmission and reflection spectra are obtained from Eq 2.

Next we study the effect of stitching error on a filter designed for a spectral bandwidth of $\Delta\lambda=0.8$nm. The coupling coefficient of $\kappa=40$ cm$^{-1}$, and the total length of $L_{TOT}=1.5$ mm are chosen to attain such bandwidth. For the analysis of channel cross talk we considered evenly spaced channels on a grid of 1.6 nm. As we show next, both lateral and longitudinal offsets affect the extinction ratio, the side-lobes, and the channel isolation. Longitudinal offsets cause a shift in the center wavelength, but have no noticeable effect on the insertion loss. The lateral offsets manifest themselves solely in the insertion loss with no effect on filter's center wavelength.

To quantify the effect of field stitches on channel isolation, we calculated the amount of cross talk between the desired channel and rest of the channels on the 1.6 nm grid. The calculation was done by integrating the reflected power spectral density of one channel over its own 3dB bandwidth and over the bandwidth of each of the other 20 channels in the grid. The ratio between the power integrated in the desired bandwidth and the highest value of the power "spilled" into another channel is considered here as the channel isolation.

### 3. Lateral Stitching Offsets

To analyze the effect of lateral stitching offsets we first calculated the power loss per stitch as a function of the stitch offset, using Finite Element Method (FEM). The results were used to construct the T-matrix of the step discontinuity. Then, per given stitch offset, the transmission and reflection spectra were calculated based on the provided model. The obtained insertion loss and the channel isolation are shown in Figs 2a and 2b, respectively. The degradation in the channel isolation is not monotonous with the number of stitches and their magnitude. This is due to the tradeoff between the insertion loss and the side-lobes. The former is increased and the latter is reduced as more stitches are introduced with an increased offset, as shown in Fig 1b. The two tendencies have an opposite effect on the channel isolation.

### 4. Longitudinal Stitching Offsets

Longitudinal stitching offset can be modeled as a section of a straight waveguide. Interestingly, such offsets cause shift in the Bragg wavelength center, $\Delta\lambda$, as shown in Fig 1d. The amount of shift ($\Delta\lambda$) was calculated as a function of the number of stitches, N, and the magnitude of the systematic longitudinal offset. The result is shown in a color-coded map in Fig 3a as a function of the two parameters, N and $\Delta x$. Our simulations show that the shift ($\Delta\lambda$), is proportional to the total offset ($N\Delta x$). Furthermore, the waveguide Bragg grating with offsets acts as if the offsets were uniformly distributed along the entire length of the device, increasing the effective period of the grating.

Another manifestation of systematic longitudinal field offset is the degradation in the channel isolation. Part of it is attributed to the shift in wavelength, and another part to the appearance of additional resonances, observed in Fig 1d in the form of the increased side-lobes. Channel isolation is shown in Fig 3b.

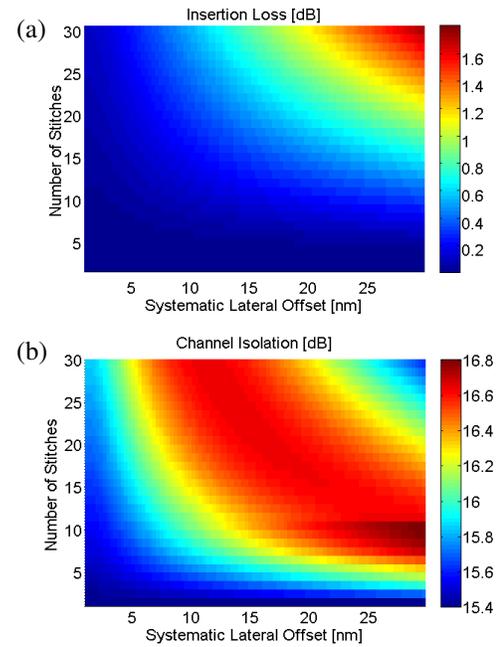

Figure 2. Analysis of systematic lateral field offset. Insertion loss (a) and channel isolation (b) as a function of number of stitches N, and the stitching error $\Delta y$ in nanometers.

## 4. Discussion

To evaluate manufacturability of long waveguide Bragg gratings with regard to the stitching errors introduced in the fabrication process we distinguish between two cases. First, consider a waveguide Bragg grating, fabricated with direct e-beam writing. Since the critical dimension of waveguide Bragg gratings is on the order of 100 nm, resolution of about 10 nm is required to assure good control over grating's profile. Existing e-beam writers achieve such resolution with write-fields as large as 500μm × 500μm. As a consequence, the entire length of the grating (1.5 mm in our case) is subdivided into several (3) write-fields, and stitching errors on the order of ~20nm or below are to be expected. For such offsets, wavelength shifts below 0.1 nm and degradation of channel isolation by <2 dB due to the longitudinal stitching errors, are anticipated.

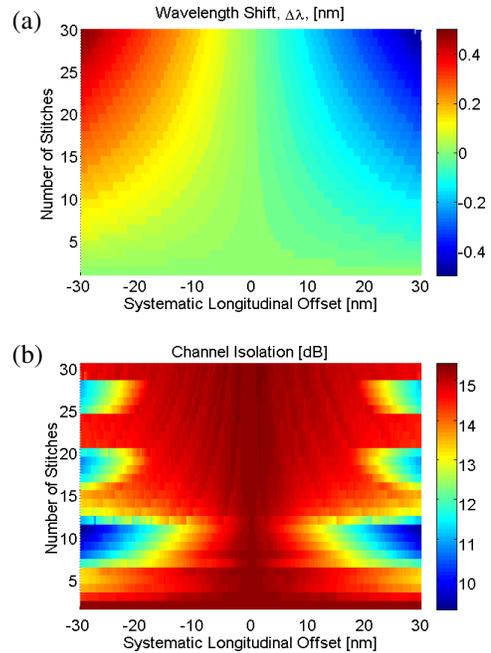

Figure 3. Analysis of systematic longitudinal field offset. Wavelength shift (a) and channel isolation (b) as a function of the number of stitches N, and the stitching error $\Delta x$ in nanometers.

Second, we consider a Bragg grating fabricated using UV-lithography. In this process the mask, created with an e-beam writer, is projected with some demagnification onto the wafer coated with a UV-sensitive resist. The exposed resist is developed and used as a mask for the following etching process, which transfers the desired profile onto the wafer. Assuming demagnification factor of M, now an M-times larger pattern needs to be written with an e-beam to create the UV mask. Assuming the same e-beam write-field as in the previous case is used to write the mask, the grating will have M-times more fields and M-times more stitches (3*M in our example), however the same offset introduced in each stitch. After the mask is projected with M-times demagnification, its image has M-times more stitches (3*M), however M-times smaller stitching error (20/M nm). To be more specific, we assume M=3 such that the device in our example will consist of ~9 stitches, with a typical offset of ~7nm. Wavelength shift due to the stitching error is still below 0.1 nm, as shown in Fig 3a, and the channel isolation is reduced by ~1.5 db, compared to an ideal Bragg grating, as shown in Fig 3b.

The significance of the effect of stitching on the performance of the Bragg grating depends on the application. For instance, dense WDM systems operate with channel spacing of 25 GHz (~0.2 nm). Wavelength shifts of ~0.1 nm, caused by the fabrication errors can therefore have a strong impact on the performance of a dense WDM optical link. Some measures can be taken to mitigate these effects, including pattern pre-distortion, optimized grating design, and minimization of the stitching errors in the e-beam lithography. Fabrication requirements can be further relaxed, by packing long gratings into a single lithographic field using a previously developed approach [7]. The obtained results provide insights into the feasibility of high-volume low-cost manufacturing of chip-scale photonic filters and multiplexers based on Bragg gratings.

This work was supported by the Defense Advanced Research Projects Agency, the National Science Foundation, the NSF CIAN ERC, and the U.S. Army Research Office. We wish to thank Sun Labs at Oracle, Nanofab staff at UCSB, Nano3 staff at UCSD, Bill Mitchell, and Rostislav Rokitski for suggestions and fruitful discussions.